# Tuning Chern Number in Quantum Anomalous Hall Insulators

Yi-Fan Zhao[1,3], Ruoxi Zhang[1,3], Ruobing Mei[1,3], Ling-Jie Zhou[1], Hemian Yi[1], Ya-Qi Zhang[1], Jiabin Yu[1], Run Xiao[1], Ke Wang[2], Nitin Samarth[1], Moses H. W. Chan[1], Chao-Xing Liu[1], and Cui-Zu Chang[1]

[1]Department of Physics, The Pennsylvania State University, University Park, PA 16802, USA

[2]Materials Research Institute, The Pennsylvania State University, University Park, PA 16802, USA

[3]These authors contributed equally: Yi-Fan Zhao, Ruoxi Zhang, Ruobing Mei

Corresponding authors: cxl56@psu.edu (C.-X. L.); cxc955@psu.edu (C.-Z. C.)

**Abstract:** The quantum anomalous Hall (QAH) state is a two-dimensional topological insulating state that has quantized Hall resistance of $h/Ce^2$ and vanishing longitudinal resistance under zero magnetic field, where $C$ is called the Chern number [1,2]. The QAH effect has been realized in magnetic topological insulators (TIs) [3-9] and magic-angle twisted bilayer graphene [10,11]. Despite considerable experimental efforts, the zero magnetic field QAH effect has so far been realized only for $C$ = 1. Here we used molecular beam epitaxy to fabricate magnetic TI multilayers and realized the QAH effect with tunable Chern number $C$ up to 5. The Chern number of these QAH insulators is tuned by varying the magnetic doping concentration or the thickness of the interior magnetic TI layers in the multilayer samples. A theoretical model is developed to understand our experimental observations and establish phase diagrams for QAH insulators with tunable Chern numbers. The realization of QAH insulators with high tunable Chern numbers facilitates the potential applications of dissipationless chiral edge currents in energy-

**efficient electronic devices and opens opportunities for developing multi-channel quantum computing and higher-capacity chiral circuit interconnects.**

**Main text:** The Chern number, also known as the Thouless-Kohmoto-Nightingale-Nijs (TKNN) number, is an integer that defines the topological phase in the quantum Hall (QH) effect [12]. This number can be calculated by the integral of the Berry curvature over the entire first Brillion zone and determines the number of topologically protected chiral edge channels along the edge of a sample [13]. The quantum anomalous Hall (QAH) effect can be considered as the QH effect under zero magnetic field which can be realized because of the intrinsic property of the electronic band structure with broken time-reversal symmetry [1,2,14,15]. The QAH effect under zero magnetic field to date has been realized in the following three systems: (*i*) molecular beam epitaxy (MBE)-grown magnetically doped topological insulator (TI) films, such as Cr- and/or V-doped $(Bi, Sb)_2Te_3$;[3,4,6-9] (*ii*) mechanically exfoliated intrinsic magnetic TI $MnBi_2Te_4$ thin flakes with odd number layers [5]; and (*iii*) manually assembled twisted bilayer graphene [10,11]. The QAH states observed in these systems were limited to $C = 1$. In $C = 1$ QAH insulators, the chiral edge current is dissipationless, but the contact resistance between the normal metal electrodes and the chiral edge channels is limited to a minimum value ($h/Ce^2$) set by Landauer theory [16,17]. This contact resistance constrains even proof-of-concept device technologies that might seek to take advantage of chiral edge channel transport at zero magnetic field. A solution is to significantly reduce this contact resistance by increasing the number of parallel chiral edge channels in QAH devices with an effective Hall resistance of $h/Ce^2$ for a high Chern number $C$.[18,19] Larger Chern number $C$ can also increase the effective breakdown current of chiral edge states and thus facilitates the practical applications of the QAH devices.

Recently, the $C = 2$ Chern insulator has been realized under a finite external magnetic field, but without well-defined Landau levels, in thin $MnBi_2Te_4$ flakes [20] and in a rhombohedral trilayer graphene/hexagonal boron nitride moiré superlattice [21]. Here, we report the realization of high Chern number QAH insulators (with $C$ tunable to integer values up to 5) with good quantization and vanishing longitudinal resistance at zero magnetic field.

The high Chern number QAH effect has been theoretically proposed in Cr-doped $Bi_2$(Se, Te)$_3$ (Ref. [18]) and magnetically doped topological crystalline insulator SnTe (Ref. [19]) films. The high Chern number QAH effect in the former system is predicted when two or more pairs of inverted sub-bands are induced by strong exchange fields [18,22], while the high Chern number QAH effect in the magnetically doped SnTe system is conjectured in the presence of multiple Dirac surface states [19]. In practice, the realization of the high Chern number QAH state in Cr-doped $Bi_2$(Se, Te)$_3$ film for strong exchange fields is unlikely, given the non-square or hysteresis-free loop and the possible metallic phase [23,24]. The high Chern number QAH state in magnetically doped SnTe is also challenging because of the absence of ferromagnetism and/or multiple Dirac points that are normally located at different energies. This property makes it difficult to have a fully gapped surface in the SnTe system [19,25]. In addition to these two systems, the high Chern number QAH state can also be realized in magnetic TI-based multilayer structures with alternating $C = 1$ QAH and normal insulator layers (Fig. 1a) [26,27]. The thickness of the normal insulator layer modulates the coupling between two $C = 1$ QAH layers and thus tunes the Chern number of QAH multilayer samples. Experimental efforts along this direction have shown the possibility of parallel connected multiple QAH layers using insulating CdSe layers. Since CdSe has the wurtzite structure, which is different from the tetradymite structure of magnetic TI, stacking faults inevitably arise in QAH/CdSe multilayer samples. These defects might be responsible for the reported large longitudinal resistance and the Hall resistances greater than the corresponding quantized values [28].

In this work, we used the MBE technique to fabricate magnetic TI/TI multilayers with symmetric structures, specifically [3 quintuple layers (QL) Cr-doped (Bi, Sb)$_2$Te$_3$ /4 QL (Bi, Sb)$_2$Te$_3$]$_m$ /3 QL Cr-doped (Bi, Sb)$_2$Te$_3$ multilayer structures, where $m$ is an integer reflecting the number of bilayer periods. We observed well-quantized QAH effect with tunable Chern number $C$ up to 5. The Chern number $C$ of the QAH insulators is determined by the number $m$ of undoped TI layers in multilayer structures. We further demonstrated that in the same multilayer sample configuration, the Chern number $C$ can be tuned by systematically varying the Cr doping concentration or by changing the thickness of the magnetic TI layers. We theoretically simulated the magnetic TI/TI multilayers with a four-band model and established phase diagrams for tunable Chern number as a function of the bulk mass parameter, spin splitting, and thickness of the magnetic TI layers. Our theoretical results are in excellent agreement with our experimental observations.

All magnetic TI/TI multilayer samples used in this work were grown on 0.25 mm thick heat-treated SrTiO$_3$(111) substrates in a MBE chamber with a base vacuum ~$2 \times 10^{-10}$ mbar. The Bi/Sb ratio in each layer is optimized to tune the chemical potential for the entire multilayer near the charge neutral point. The electrical transport measurements were carried out in a Physical Property Measurement System (Quantum Design, 2 K, 9 T) and in a dilution refrigerator (Leiden Cryogenics, 10 mK, 9 T) cryostat with the magnetic field applied perpendicular to the film plane. Six-terminal mechanically defined Hall bars with bottom-gate electrodes were used for electrical transport studies (see Methods for details).

Similar to the $C = 1$ QAH insulators [3-10], the high Chern number QAH insulator harbors $C$ dissipationless chiral edge channels [18,19]. These edge states are spin-polarized, and their chirality is determined by the internal magnetization of the sample (Figs. 1b). As noted above, the high Chern number QAH insulators can be realized in the $C = 1$ QAH/normal insulator multilayer structures [26]. The heavily Cr-doped (Bi, Sb)$_2$Te$_3$ layer (i.e. $x = 0.24$ in (Bi, Sb)$_{2-x}$Cr$_x$Te$_3$) in our magnetic TI/TI multilayer structures plays a two-fold role: (*i*) its magnetism breaks the time-reversal symmetry of the 4QL (Bi, Sb)$_2$Te$_3$ layer and thus allows for the $C = 1$ QAH effect to emerge in the 3QL (Bi, Sb)$_{1.76}$Cr$_{0.24}$Te$_3$ /4QL (Bi, Sb)$_2$Te$_3$/3QL (Bi, Sb)$_{1.76}$Cr$_{0.24}$Te$_3$ sandwich (Fig. 2a); (*ii*) the bulk band gap of the magnetic TI layer tends to become non-inverted because the heavy Cr doping greatly reduces the spin-orbit coupling (SOC) of the system [23,24]. In addition, the Chern number $C$ of the magnetic TI/TI multilayer samples can be determined by the strength of the coupling between two adjacent $C = 1$ QAH layers, i.e. the thickness of the interior magnetic TI layers. When this coupling is weak, i.e. the magnetic TI layer is thick, the high Chern number QAH insulators emerge. When this interaction is stronger than a critical value, i.e. the magnetic TI is thin enough or even absent, the sample shows the QAH insulator state with a Chern number $C = 1$. Figure 1c shows the cross-sectional scanning transmission electron microscopy (STEM) image of the $m = 2$ multilayer sample (i.e. the (3 QL (Bi, Sb)$_{1.76}$Cr$_{0.24}$Te$_3$ /4 QL (Bi, Sb)$_2$Te$_3$)$_2$ /3QL (Bi, Sb)$_{1.76}$Cr$_{0.24}$Te$_3$ sample) grown on SrTiO$_3$ substrate. Since the magnetically doped TI and undoped TI shares the same lattice structure, our samples show a highly-ordered lattice structure (Extended Data Figs. 1 and 2), which, we will show in detail below, is of great importance for the realization of the tunable high Chern number QAH effect with good quantization.

We next carried out magneto-transport measurements on the [3QL (Bi, Sb)$_{1.76}$Cr$_{0.24}$Te$_3$ /4QL (Bi, Sb)$_2$Te$_3$]$_m$ /3QL (Bi, Sb)$_{1.76}$Cr$_{0.24}$Te$_3$ multilayer samples with $1 \leqslant m \leqslant 5$ at $T = 25$ mK when the bottom gate voltage $V_g$ is tuned at the charge neutral point $V_g^0$ (Fig. 2). All five samples exhibit the QAH effect with varying degrees of quantization presicions and longitudinal resistance minima. For the $m = 1$ sample, the Hall resistance $\rho_{yx}$ displays the quantized value of 0.994 $h/e^2$, and the longitudinal resistance $\rho_{xx}$ is 0.0001 $h/e^2$ (~ 2.5 Ω) under zero magnetic field, corresponding to the QAH state with Chern number $C = 1$ (Fig. 2a). By inserting one more period of the 3QL (Bi, Sb)$_{1.76}$Cr$_{0.24}$Te$_3$/4QL (Bi, Sb)$_2$Te$_3$ bilayer, $\rho_{yx}$ displays the quantized value of 0.498 $h/e^2$, and $\rho_{xx}$ is 0.008 $h/e^2$ under zero magnetic field for the $m = 2$ sample, giving rise to the QAH state with Chern number $C = 2$ (Fig. 2b). Upon systematically inserting more periods of the 3QL (Bi, Sb)$_{1.76}$Cr$_{0.24}$Te$_3$/4QL (Bi,Sb)$_2$Te$_3$ bilayers, $\rho_{yx}$ displays the quantized values of 0.329$h/e^2$, 0.234 $h/e^2$, and 0.185 $h/e^2$ for the $m = 3$, 4, and 5 samples, respectively. The corresponding $\rho_{xx}$ in these three samples are respectively 0.022$h/e^2$, 0.039 $h/e^2$, and 0.044 $h/e^2$ under zero magnetic field (Figs. 2c to 2e). These are the QAH states with Chern number $C = 3$, 4, and 5. We speculate that the increasing value of $\rho_{xx}$ under zero magnetic field with increasing number of the periods in the multilayer samples is likely due to the following three sources: (*i*) Conduction from the dissipative quasihelical side surface states increases in thicker samples [17,29]; (*ii*) Conduction from the dissipative residual bulk carriers increases in thicker samples; and (*iii*) Tuning the chemical potentials of two surfaces simultaneously into magnetic exchange gaps becomes harder in thicker QAH samples via a single bottom gate $V_g$.[30] Further studies are needed to clarify which source is dominant in these thick TI/magnetic TI multilayer samples.

The realization of the high Chern number QAH insulators with $C$ of 1 to 5 in magnetic TI/TI multilayer samples is further validated by the gate $V_g$ dependence of $\rho_{yx}$ and $\rho_{xx}$ at zero magnetic field (labeled as $\rho_{yx}(0)$ and $\rho_{xx}(0)$), respectively (Fig. 3). $\rho_{yx}(0)$ in all five samples exhibit distinct plateaus with the quantized values $h/Ce^2$, all of which are centered at the charge neutral point $V_g = V_g^0$. Accompanying the quantization in $\rho_{yx}(0)$, the zero magnetic field longitudinal resistance $\rho_{xx}(0)$ are greatly suppressed for the $m = 1$ to 5 samples, respectively (Figs. 3a to 3e). The ratios $\rho_{yx}(0)/\rho_{xx}(0)$ correspond to anomalous Hall angles of 89.99º, 89.08º, 86.17º, 80.54º, and 76.55º for the $m = 1$ to 5 samples, respectively. Here, we define the critical temperature of the high Chern number QAH state as that at which the $\rho_{yx}(0)/\rho_{xx}(0)$ ratio is greater than 1 (i.e. the crossing point between $\rho_{yx}(0)$-$T$ and) $\rho_{xx}(0)$-$T$ curves in Extended Data Figs. 3 to 7). The critical temperatures are 3.5 K, 2.6 K, 8.3 K, 6.7 K, and 6.3 K for the $C = 1$ to 5 samples, respectively.

The realization of the perfect high Chern number QAH insulator by alternatively stacking the $C = 1$ QAH and normal insulator layers (Fig. 1a) is constrained by the inevitable presence of quasihelical side surface conductions in thick multilayer samples arising from strong coupling between two adjunct $C = 1$ QAH layers [17,29]. Therefore, dissipationless transport under zero magnetic field (i.e. $\rho_{xx}(0) = 0$) of stacked QAH layers cannot be realized. However, if the coupling between two $C = 1$ QAH layers can be greatly suppressed by increasing the thickness of the normal insulator layer, the side surfaces from all QAH layers cannot be connected, which can prevent the formation of additional quasihelical modes, and therefore the side surface conduction contribution can be greatly suppressed. Therefore, in principle, it is feasible to synthesize the perfect high Chern number QAH insulators with $\rho_{xx}(0)$ = 0 via the stacking approach.

In the following, we showed that the Chern numbers of these QAH insulators can be tuned by controlling either the magnetic doping concentration or the thickness of the interior magnetic TI layers. We first systematically changed the Cr doping concentration $x$ in the $m = 2$ sample (i.e. [3QL (Bi, Sb)$_{2-x}$Cr$_x$Te$_3$ /4QL (Bi, Sb)$_2$Te$_3$]$_2$ /3QL (Bi, Sb)$_{2-x}$Cr$_x$Te$_3$). Like the $m = 2$ sample with $x = 0.24$ shown in Figs. 2b and 3b, the sample with higher Cr doping concentrations (i.e. $x = 0.35$) also shows the $C = 2$ QAH effect with the zero magnetic field Hall resistance $\rho_{yx}(0)$ of 0.497 $h/e^2$. In contrast, the $m = 2$ samples with lower Cr doping concentration (i.e. $x = 0.13$ and $x = 0.15$) show the $C = 1$ QAH effect with the zero magnetic field Hall resistance $\rho_{yx}(0)$ of 0.969 $h/e^2$ for the $x = 0.13$ sample and 0.996 $h/e^2$ for the $x = 0.15$ sample (Fig. 4a). Such a change of the Chern number $C$ from 2 to 1 occurs in the $m = 2$ samples because lowering the Cr concentration not only drives the bulk energy gap of the magnetic TI layers towards the inverted (i.e. nontrivial) regime [23,24] but also reduces their exchange spin splitting [2,13,14]. As a result, nontrivial interface states disappear at the interfaces between the interior magnetic TI and undoped TI layers with lowering Cr concentration, so only a pair of nontrivial interface states localized at the top and bottom magnetic TI layers of the $m = 2$ sample contribute to the total Hall resistance (i.e. $C = 1$) (Fig. 4c). For the higher Cr doping concentrations, the SOC of the magnetic TI layer is weakened and it tends to become a trivial insulator layer, while the exchange spin splitting is enhanced. Consequently, two nontrivial interface states are formed at the interface between the interior magnetic TI and TI layers, resulting in two pairs of nontrivial interface states contributing to the total Hall resistance (i.e. $C = 2$) (Fig. 4c).

Therefore, the Chern number of the QAH insulators in the magnetic TI/TI multilayer samples can be tuned by controlling the magnetic doping concentration of the magnetic TI layers.

We next changed the thickness $d$ of the middle magnetic TI layer in the $m = 2$ sample (i.e. [3QL (Bi, Sb)$_{1.76}$Cr$_{0.24}$Te$_3$ /4QL (Bi, Sb)$_2$Te$_3$]/$d$ QL (Bi, Sb)$_{1.76}$Cr$_{0.24}$Te$_3$/[4QL (Bi, Sb)$_2$Te$_3$ /3QL (Bi, Sb)$_{1.76}$Cr$_{0.24}$Te$_3$]), which can modulate the coupling between two $C = 1$ QAH layers. For the $d = 0$ and 1 QL samples, the coupling between the top and bottom two QAH layers is strong and only a pair of nontrivial interface states exist (Fig. 4c), so these two samples show the $C = 1$ QAH effect. Indeed, $\rho_{yx}(0)$ is quantized at $0.995h/e^2$ for the $d = 0$ sample and $0.996h/e^2$ for the $d = 1$ sample, respectively (Fig. 4b). For the $d \geq 2$ QL samples, the interaction between the top and bottom QAH layers is reduced and one more pair of the nontrivial interface states emerges (Fig. 4c), resulting in the $C = 2$ QAH effect in these three samples. $\rho_{yx}(0)$ is quantized at $0.469h/e^2$ for the $d = 2$ sample and $0.491h/e^2$ for the $d = 4$ sample (Fig. 4b). Therefore, the Chern number of the QAH insulators in the magnetic TI/TI multilayers can also be tuned by adjusting the thickness $d$ of the interior magnetic TI layer.

To buttress our interpretation, we numerically simulated the magnetic TI/TI multilayer structures based on a four-band model [31] in the basis $|P1_-^+,\uparrow\rangle$, $|P2_+^-,\uparrow\rangle$, $|P1_-^+,\downarrow\rangle$ and $|P2_+^-,\downarrow\rangle$ with the Hamiltonian

$$H(\boldsymbol{k}_\parallel, -i\partial_z) = \epsilon(\boldsymbol{k}_\parallel, -i\partial_z) + M(\boldsymbol{k}_\parallel, -i\partial_z)\tau_z + B(-i\partial_z)\tau_y + A(k_y\tau_x\sigma_x - k_x\tau_x\sigma_y)$$
$$+ g(z)\sigma_z,$$

where $\epsilon(\boldsymbol{k}_\parallel, -i\partial_z) = C_0 + C_1(-\partial_z^2) + C_2 k_\parallel^2$, and $M(\boldsymbol{k}_\parallel, -i\partial_z) = \mathcal{M}_0(z) + M_1(-\partial_z^2) + M_2 k_\parallel^2$ with $\boldsymbol{k}_\parallel = (k_x, k_y)$ and the material-dependent parameters $C_0, C_1, C_2, \mathcal{M}_0, M_1, M_2, A, B, g$. The parameter $\mathcal{M}_0$, dubbed the mass parameter here, is the key parameter that characterizes the inverted ($\mathcal{M}_0 < 0$) and non-inverted ($\mathcal{M}_0 > 0$) band structures [32]. The Pauli matrices $\boldsymbol{\sigma}$ and $\boldsymbol{\tau}$ represent spin and orbital degrees of freedom. To simulate the multilayer structures with the total multilayer thickness $L$, we considered the eigen-equation $H(\boldsymbol{k}_\parallel, -i\partial_z)\psi_{\boldsymbol{k}_\parallel}(z) = E\psi_{\boldsymbol{k}_\parallel}(z)$ with the open boundary condition $\psi_{\boldsymbol{k}_\parallel}(z = 0) = \psi_{\boldsymbol{k}_\parallel}(z = L) = 0$. We noted that the Cr doping in the magnetic TI layers plays two roles: (*i*) it reduces the SOC of the magnetic TI layers and drives their bulk energy gap towards the normal insulator regime (described by the mass parameter $\mathcal{M}_0$) [23] and (*ii*) it introduces magnetic moments in TI, resulting in a spin splitting (described by the $g$ term) due to the exchange coupling [2]. Consequently, we considered the $z$ dependence of these parameters as $\mathcal{M}_0(z), g(z) = \begin{cases} M_0, g, & z \in \text{Cr doped TI layer} \\ M_0', 0, & z \in \text{undoped TI layer} \end{cases}$. We numerically solve the eigen-equation and compute the Hall conductance $\sigma_{xy}$ with the numerical methods described in Methods Section. In the following, we focused on the $m = 2$ multilayer structure (i.e. $L = 17$ nm) to interpret the tunable Chern number observed in our experiments (Figs. 4a and 4b).

We first studied the process of changing the Cr concentration. Since the Cr dopants primarily affect the mass parameter $M_0$ and the exchange coupling term $g$ in the magnetic TI layers, we plotted the bulk energy gap of the $m = 2$ multilayer sample as a function of $M_0$ and $g$ in Fig. 4d. The energy gap closing is depicted by the bright yellow lines which separate different phases. The process of adding Cr dopants in our experiments qualitatively corresponds to moving from point I to point V along the dashed line shown in Fig. 4d, which shows appropriately that increasing Cr concentration can increase both $M_0$ and $g$. Our calculation suggests that the Hall conductance $\sigma_{xy}$ takes the value of $\frac{e^2}{h}$ for points I, II, III (i.e. $C = 1$) and $\frac{2e^2}{h}$ for points IV, V (i.e. $C = 2$), thus qualitatively reproducing the influence of Cr doping in our experiments (Fig. 4a).

Our simulation provides more insights into the physical mechanism. We found that the Hall conductance $\sigma_{xy}$ predominantly comes from the interface Dirac bands, i.e., the bands that have dispersions like gapped Dirac fermions and are spatially localized at the interfaces between magnetic TI and undoped TI layers (Extended Data Fig. 15). For the $m$ = 2 sample, there are four interfaces, two outer ones and two inner ones (Fig. 4c). In the $C$ = 2 phase, we have one occupied interface Dirac band on each of the four interfaces; each interface Dirac band contributes $\frac{e^2}{2h}$ to the total Hall conductance $\sigma_{xy}$ and thus four Dirac bands add up to $\sigma_{xy} = \frac{2e^2}{h}$ (i.e. $C$ = 2), as shown in Fig. 4c and Extended Data Fig. 15 and further supported by our calculations in Fig. 4f. Upon reducing the Cr doping concentration in the magnetic TI layer, two sets of Dirac bands at the two inner interfaces penetrate more into the bulk, leading to a gap closing and re-opening with a trivial hybridization gap. After this topological phase transition, the Hall conductance $\sigma_{xy}$ comes only from the Dirac bands on the two outer interfaces, as shown in Fig. 4c and Extended Data Fig. 16.

Next, we turned to the process of reducing the thickness $d$ of the middle magnetic TI layer in the $m$ = 2 multilayer sample. As shown in Fig. 4e, reducing $d$ can also induce the topological phase transition from $C$ = 2 to $C$ = 1 QAH insulators. With appropriate values of $M_0$ and $g$, our calculation indicates that the critical thickness value for the transition lies between 1 nm and 2 nm (corresponding to the one and two QLs), coinciding with our experimental observations on the change of the Chern number $C$ in Fig. 4b.

Our calculations further show that the $C$ = 2 phase can also occur at negative $M_0$ (i.e. the nontrivial regime of the magnetic TI layer) once $g$ is sufficiently large, but increasing $M_0$ can significantly expand the $C = 2$ phase region (Fig. 4d). For even larger $g$, an additional gap-closing line separates the QAH insulator phases with $C = 2$ and $C = 3$ (Fig. 4d). However, since doping more Cr atoms in TI samples will simultaneously increase $M_0$, the $C = 3$ phase is unlikely to be achieved for the $m = 2$ multilayer structure in real experiments.

To summarize, we fabricated magnetic TI/TI multilayer structures and realized QAH insulators with Chern number $C$ from 1 to 5 under zero magnetic field. The Chern number in the same sample configuration can be tuned by varying either the magnetic doping concentration or the thickness of the interior magnetic TI layers. The realization of QAH insulators with tunable Chern number at zero magnetic field illuminates and expands the known topological phases of quantum matter. This advance also provides an important platform for demonstrating proof-of-concept applications that exploit the dissipation-free chiral edge current of QAH insulators. For instance, the greatly reduced contact resistance in high Chern number QAH devices allows information to be multiplexed over multiple chiral edge channels and thus is of interest for higher-capacity chiral circuit interconnects. The realization of tuning Chern number in QAH insulators also provides a model system for studying topological quantum phase transitions with time-reversal symmetry breaking and introduces the channel degree of freedom of chiral edge states, thus opening the possibility of using topologically protected and dissipation-free chiral edge states to store and transfer quantum information. Finally, magnetic TI/TI multilayer samples could serve as a rich platform for discovering and studying many other emergent topological phenomena, such as time-reversal symmetry breaking Weyl semimetal phases with only one pair of Weyl nodes [26] and dynamical axion electrodynamics [33,34].

**Methods**

**Growth of magnetic TI/TI multilayer structures**

Magnetic TI/TI multilayer samples were fabricated in a commercial MBE system (Scienta Omicron Lab10) with a base pressure lower than $2 \times 10^{-10}$ mbar. The insulating 0.25 mm $SrTiO_3$ (111) substrates used for growth were first soaked in hot deionized water (~ 80 °C) for 1.5 hours, and then annealed at 982 °C for 3 hours in a tube furnace with flowing oxygen. Through this annealing process, the surface of $SrTiO_3$ substrates was passivated and atomically flat, which are suitable for the MBE growth of TI films. The heat-treated $SrTiO_3$ (111) substrates were next loaded into the MBE chamber and outgassed at 600 °C for 1 hour before the growth of magnetic TI/TI multilayer samples. High-purity Bi(99.999%), Sb(99.9999%), Cr(99.999%), and Te(99.9999%) were evaporated from Knudsen effusion cells. During the growth of the samples, the substrate was maintained at ~ 230 °C. The flux ratio of Te per (Bi +Sb + Cr) was set to greater than 10 to prevent Te deficiency in the films. The growth rate for the films was ~0.2 QL per minute. Epitaxial growth was monitored by *in-situ* reflection high energy electron diffraction (RHEED) patterns, where the high crystal quality and the atomically flat surface were confirmed by the streaky and sharp "1×1" patterns (Extended Data Fig. 1). No capping layer is involved in the *ex-situ* electrical transport measurements.

**Hall-bar device fabrication.**

Magnetic TI/TI multilayer samples on 2 mm × 10 mm $SrTiO_3$ (111) substrates were scratched into a Hall bar geometry using a computer-controlled probe station. The effective area of the Hall bar device is ~ 1 mm × 0.5 mm (Extended Data Fig. 3a inset). The electrical ohmic-contacts for transport measurements were made by pressing indium dots on the Hall bar. The bottom gate electrode was prepared by flattening indium spheres on the backside of the SrTiO$_3$ substrate.

**Electrical transport measurements**

Transport measurements were conducted using both a Quantum Design Physical Property Measurement System (2 K, 9 T) and a Leiden Cryogenics dilution refrigerator (10 mK, 9 T) with the magnetic field applied perpendicular to the film plane. The bottom gate voltage $V_\text{g}$ was applied using the Keithley 2450. The excitation currents used in the PPMS ($\geq$ 2 K) and dilution (< 2 K) measurements are 1 μA and 1 nA, respectively. The results reported here have been reproduced on at least three samples for each Chern number (Extended Data Fig. 10). All magneto-transport results shown here were symmetrized or anti-symmetrized as a function of the magnetic field to eliminate the effect of electrode misalignments. The raw data of Fig. 2 are shown in Extended Data Fig. 8. More transport results are found in Extended Data Figs. 3 to 12.

**Theoretical simulations and calculations**

To numerically solve the eigen-equation for magnetic TI/TI multilayer structures, we expanded the wavefunction of the system as $\psi_{k_\parallel}(z) = \frac{1}{\mathcal{N}} \sum_{n,\lambda} a_{n,\lambda}(k_\parallel) \sqrt{\frac{2}{L}} \sin(\frac{n\pi}{L} z)|\lambda\rangle$ with the basis function $\sqrt{\frac{2}{L}} \sin(\frac{n\pi}{L} z)|\lambda\rangle$ where $\sin(\frac{n\pi}{L} z)$ satisfies the open boundary condition [i.e. $\psi_{k_\parallel}(z=0) = \psi_{k_\parallel}(z=L) = 0$] and $|\lambda\rangle$ is the basis of the four-band model, $a_{n,\lambda}(k_\parallel)$ is the expansion coefficient, $\mathcal{N}$ is the overall normalization fact, and $n = 1, \cdots, N$. From the eigen-equation and the expansion form, we obtained $\sum_{n',\lambda'} a_{n',\lambda'} \langle n,\lambda|H|n',\lambda'\rangle = E a_{n,\lambda}$, in which specifically each matrix element is $\langle n,\lambda|H|n',\lambda'\rangle = \frac{2}{L} \int_0^L dz \sin\left(\frac{n\pi}{L} z\right) H_{\lambda\lambda'}(z) \sin\left(\frac{n'\pi}{L} z\right)$. The parameters $\mathcal{M}_0, g$ in the Hamiltonian matrix element $H_{\lambda\lambda'}$ take different values for the Cr-doped TI and undoped TI layers and thus $H_{\lambda\lambda'}$ is z-dependent. In our calculations, we choose $\epsilon = 0$, $M_0' = -0.28\, eV$, $M_1 = 6.86\, eV\text{Å}^2$, $M_2 = 44.5\, eV\text{Å}^2$, $A = 3.33\, eV\text{Å}$, $B = 2.26\, eV\text{Å}$ and $N = 50$. All these parameters are for Bi$_2$Se$_3$ from Table IV of Ref. [31]. We have also done the same calculations using the parameters for Sb$_2$Te$_3$ (Extended Data Fig. 13) and found the phase diagrams are qualitatively similar to the ones shown in Figs. 4d and 4e.

With the above matrix elements, we can plot the energy dispersion of the magnetic TI/TI multilayer samples. We first studied an individual TI thin film with $L = 5\,nm$ and $M_0 = -0.28eV$ to gain more insight into this model [31]. For $g = 0\, eV$, i.e. the TI film without Cr doping, each band is doubly degenerate because of the spin degeneracy and there are two Dirac surface states near $\Gamma$ point (Extended Data Fig. 14a). For $g = 0.02\, eV$, i.e. the TI film with Cr doping, the non-zero magnetization term appears and induces the spin splitting. As a result, each double-degenerate band splits into two subbands (Extended Data Fig. 14b). Meanwhile, the two Dirac surface states are gapped, which gives rise to a total Chern number $C$ of one, with each gapped surface state contributing $C = 1/2$. Therefore, the total Hall conductance $\sigma_{xy}$ of the individual magnetic TI film is $\frac{e^2}{h}$.

Next, we applied our model to the *m* = 2 magnetic TI/TI multilayer structure. We set the thickness of magnetic TI and undoped TI layers to be 3 nm and 4 nm, respectively. First, we considered the magnetic TI layers with a normal insulating gap (i.e. $M_0 > 0$) and found the low energy physics is mainly determined by the undoped TI layers in the multilayer structure (Extended Data Fig. 15a). When $M_0 = 0.5\, eV$, for $g = 0\, eV$ (i.e. without magnetization in magnetic TI layer) each band is four-fold degenerate due to the existence of the two undoped TI layers in the *m* = 2 sample. When $g$ increases to $0.2\, eV$ (i.e. with magnetization in magnetic TI layer) the spin splitting occurs and thus the surface states are gapped because of the appearance of the magnetic proximity effect (Extended Data Fig. 15b). Therefore, four gapped Dirac states contribute to a total Chern number $C$ of two.

In the main text, we argued that the wavefunctions of these gapped Dirac bands are primarily located at the interfaces between magnetic TI and undoped TI layers. We here plotted the wavefunctions in the $m = 2$ magnetic TI/TI multilayer sample along $z$ direction ($0 \leq z \leq 17$ nm) for the band dispersion at $g = 0.2\ eV$ and $M_0 = 0.2\ eV$ (Extended Data Fig. 15c). Extended Data Figs. 15d to 15g show the distribution of probability for the four lowest energy conduction bands at $\Gamma$ point as a function of $z$, in which the red vertical dashed lines represent the interfaces between the Cr-doped TI and undoped TI layers. We found that the wavefunctions of the 1st band and the 4th band are mostly distributed at the two inner magnetic TI/TI interfaces of the $m = 2$ sample, while the wavefunctions of the 2nd and the 3rd bands are located at the two outer magnetic TI/TI interfaces and these two bands are almost degenerate in energy.

We also found that the $C = 2$ phase can be tuned to the $C = 1$ phase by reducing either $M_0$ or $g$. We examined this topological phase transition by fixing $g = 0.2eV$ and draw the magnetic exchange gap size as a function of the bulk gap of the magnetic TI layer $M_0$. Extended Data Fig. 16a demonstrates that a gap-closing occurs at $M_0 = -0.12\ eV$, indicating the occurrence of a topological phase transition. This topological phase transition occurs at the two inner magnetic TI/TI interfaces and drives the whole system into the $C = 1$ phase. To validate this statement, we plotted the wavefunction distribution of the four lowest energy conduction bands in Extended Data Fig. 16b, at $g = 0.2\ eV$ and $M_0 = -0.15\ eV$ (Extended

Data Figs. 16c to 16f). We found that the 1st and 4th bands, which are localized at two inner magnetic TI/TI interfaces for the $C$ = 2 phase, penetrate deeply into the middle Cr-doped TI layer and hybridize strongly with each other (Extended Data Figs. 16c and 16f) so that they should be treated as the bulk states. This strong hybridization induces the topological phase transition between the $C$ = 2 and $C$ = 1 phases. In contrast, the 2nd and 3rd bands remain spatially well separated at two outer magnetic TI/TI interfaces while their wavefunctions extend more to the two outer Cr-doped TI layers (Extended Data Figs. 16d and 16e).

To confirm the topological property of the system, we further computed the Hall conductance $\sigma_{xy}$ of a few points shown in Figs. 4d and 4e with the expression $\sigma_{xy} = \frac{e^2}{h}\sum_{n\in occupied} C_{xy}(n)$ based on the Kubo formula for the Chern number:

$$C_{xy}(n) = \frac{1}{2\pi}\int dk_x dk_y \sum_m \frac{-i\left(\langle u_n|\partial_x H|u_m\rangle\langle u_m|\partial_y H|u_n\rangle - \langle u_n|\partial_y H|u_m\rangle\langle u_m|\partial_x H|u_n\rangle\right)}{(\varepsilon_n - \varepsilon_m)^2}$$

where $\partial_x H = \frac{\partial H}{\partial k_x}$, $n$ and $m$ denote the occupied band number, and $\varepsilon_n, u_n$ are the eigenvalue and eigenvector (corresponding to the periodic part of the Bloch state) for the $n^{th}$ band.

**Acknowledgments:** We are grateful to Y. T. Cui, J. Jain, W. D. Wu, D. Xiao, and X. D. Xu for helpful discussion. The sample synthesis, dilution transport measurements, and the theoretical calculations are supported by the DOE grant (DE-SC0019064). C. Z. C. acknowledges support from the ARO Young Investigator Program Award (W911NF1810198), the NSF-CAREER award (DMR-1847811), and the Gordon and Betty Moore Foundation's EPiQS Initiative (GBMF9063 to C. Z. C.). C.X. L. acknowledges support from the ONR grant (No. N00014-18-1-2793) and Kaufman New Initiative research grant KA2018-98553 of the Pittsburgh Foundation. M. H. W. C. acknowledges support from NSF grant DMR-1707340. N. S. and R.

X. acknowledge support from the DOE EFRC grant (DE-SC0019331).

**Author contributions:** C. Z. C. conceived and designed the experiment. Y. F. Z., L. J. Z., and Y. Q. Z. grew the magnetic TI/TI multilayer samples and carried out the PPMS transport measurements with the help of C. Z. C. R. Z., L. J. Z., and Y.Q. Z. carried out the dilution transport measurements with the help of M. H. W. C. and C. Z. C. R. M., J. Y., and C.X. L. did all calculations and provided theoretical support. Y. F. Z., R. Z., R. M., C. X. L., and C. Z. C. analyzed the data and wrote the manuscript with inputs from all authors.

**Competing interests:** The authors declare no competing interests.

**Data availability:** The datasets generated during and/or analyzed during this study are available from the corresponding author upon reasonable request.

**Figures and figure captions:**

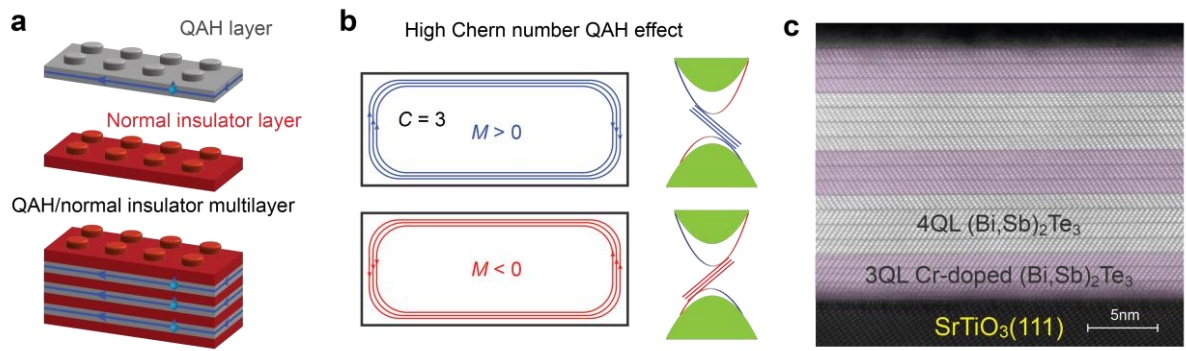

**Fig. 1| High Chern number QAH effect in $C = 1$ QAH/normal insulator multilayer structures. a**, Schematics (depicted as stacking Lego layers) of the high Chern number QAH insulator in $C = 1$ QAH/normal insulator multilayer structures. **b**, Schematics of the high Chern number QAH effect. We here take $C = 3$ QAH insulator as an example. Top: three chiral edge channels in the real space (left) and momentum space (right) for the magnetization $M > 0$. Bottom: three chiral edge channels in the real space (left) and momentum space (right) for the magnetization $M < 0$. **c**, Cross-sectional scanning transmission electron microscopy (STEM) image of the $m = 2$ magnetic TI/TI sample grown on SrTiO$_3$ substrate.

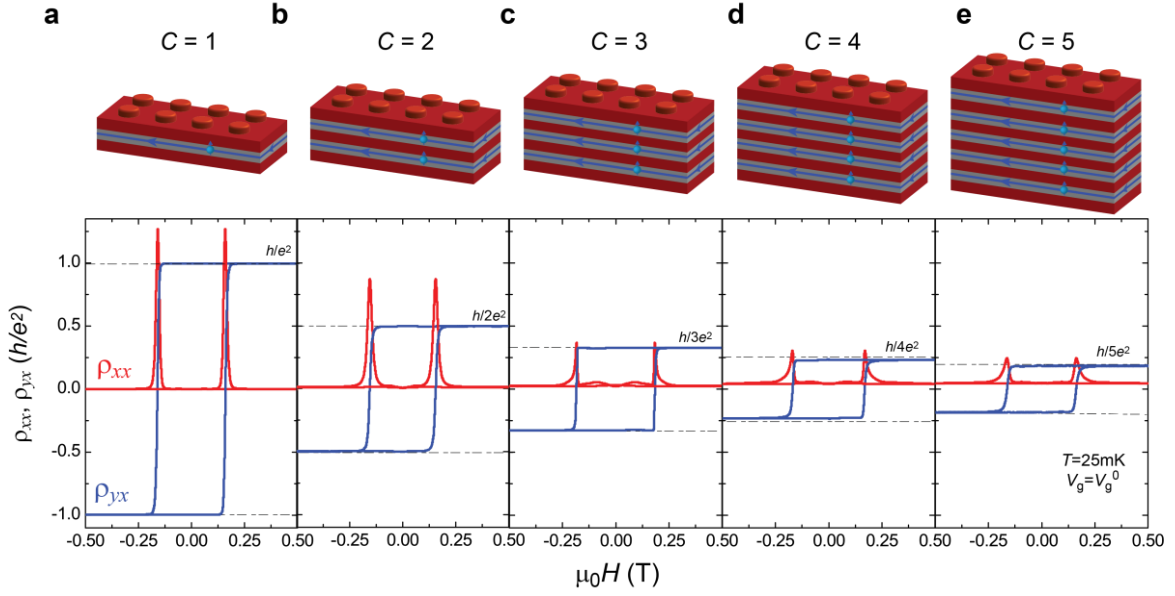

**Fig. 2| High Chern number QAH effect observed in magnetic TI/TI multilayer samples.**

**a-e**, Top: schematics (Legos) of the multilayer structures for the QAH effect with Chern number from $C = 1$ to $C = 5$. Bottom: magnetic field $\mu_0 H$ dependence of the longitudinal resistance $\rho_{xx}$ (red curve) and the Hall resistance $\rho_{yx}$ (blue curve) measured at the charge neutral point $V_g = V_g^0$ and $T = 25$ mK. $V_g^0$ of these five samples are +11 V ($C = 1$), +10 V ($C = 2$), +2 V ($C = 3$), +2 V ($C = 4$), and -3 V ($C = 5$). The thicknesses of these five samples are 10 nm ($C = 1$), 17 nm ($C = 2$), 24 nm ($C = 3$), 31 nm ($C = 4$), and 38 nm ($C = 5$).

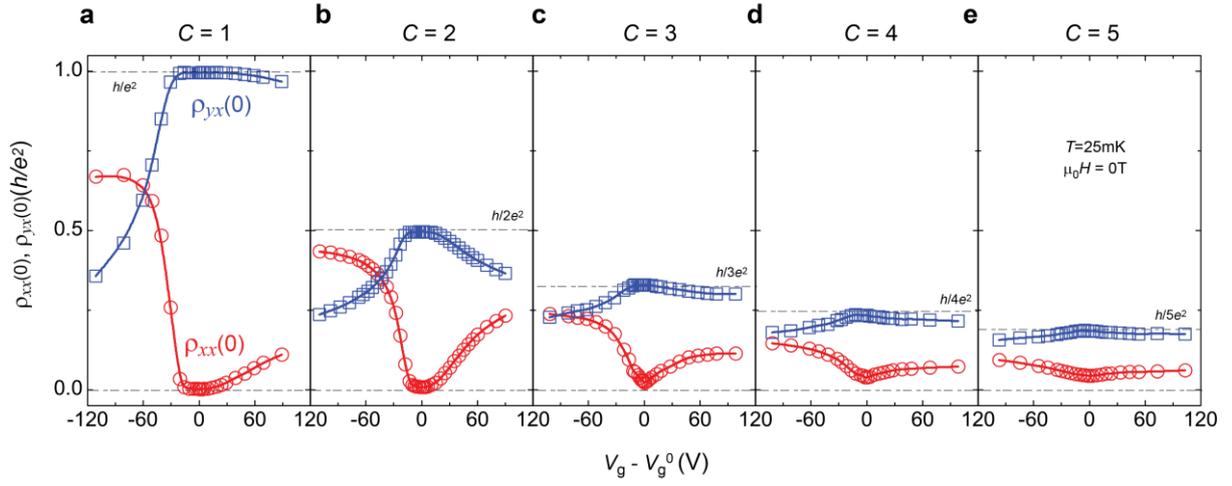

**Fig. 3| Demonstration of high Chern number QAH state in magnetic TI/TI multilayer samples. a-e**, Gate ($V_g$-$V_g^0$) dependence of $\rho_{yx}(0)$ (blue squares) and $\rho_{xx}(0)$ (red circles) of the QAH insulators with Chern number $C$ of 1 to 5. All measurements were taken at 25 mK and $\mu_0 H = 0$ T after magnetic training.

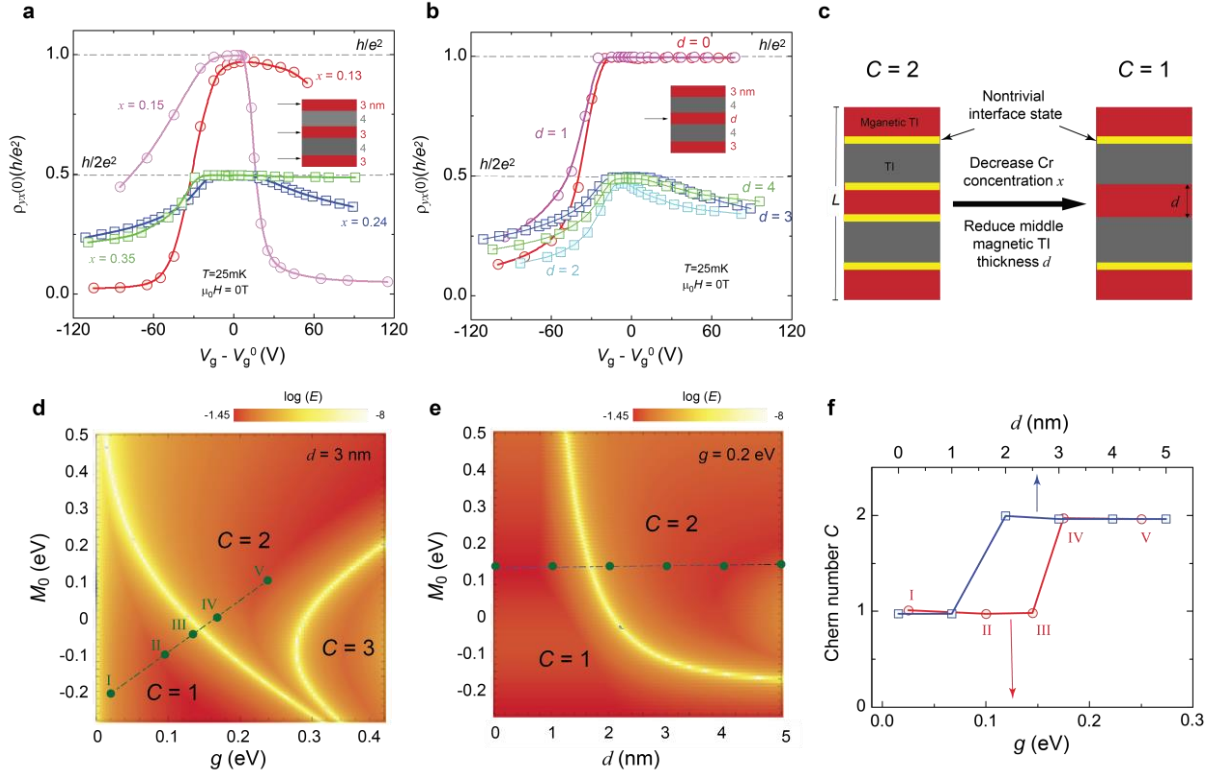

**Fig. 4| Tunable Chern number in QAH insulators. a**, Gate ($V_g$-$V_g^0$) dependence of the zero magnetic field Hall resistance $\rho_{yx}(0)$ for the $m$ =2 samples (i.e. [3QL (Bi, Sb)$_{2-x}$Cr$_x$Te$_3$ /4QL (Bi, Sb)$_2$Te$_3$]$_2$ /3QL (Bi, Sb)$_{2-x}$Cr$_x$Te$_3$) with different Cr doping level $x$. The $x$ = 0.13 and 0.15 samples show the $C$ = 1 QAH effect, while the $x$ = 0.24 and 0.35 samples show the $C$ = 2 QAH effect. $V_g^0$ of these four samples are +65 V ($x$ = 0.13), -10 V ($x$ = 0.15), +10 V ($x$ = 0.24), and +9 V ($x$ = 0.35). **b**, Gate ($V_g$-$V_g^0$) dependence of $\rho_{yx}(0)$ for the $m$ =2 samples (i.e. [3QL (Bi, Sb)$_{1.76}$Cr$_{0.24}$Te$_3$ /4QL (Bi, Sb)$_2$Te$_3$]/$d$ QL (Bi, Sb)$_{1.76}$Cr$_{0.24}$Te$_3$/[4QL (Bi, Sb)$_2$Te$_3$ /3QL (Bi, Sb)$_{1.76}$Cr$_{0.24}$Te$_3$]) with varying the thickness of the middle magnetic TI layer $d$. The $d$ = 0 and 1 samples show the $C$ = 1 QAH effect, while the $d$ = 2 to 4 samples show the $C$ = 2 QAH effect. $V_g^0$ of these four samples are +20 V ($d$ =0 ), +23 V ($d$ =1 ), +18 V ($d$ =2 ), +10 V ($d$ =3), and +4 V ($d$ =4). **c**, Schematics for nontrivial interface states (marked by the yellow color) in the $m$ =2 sample, which can be tuned by controlling the Cr concentration $x$ and the middle magnetic

TI thickness $d$. A pair of nontrivial interface states adds one to the total Chern number $C$. **d**, The calculated phase diagram for the $m = 2$ sample as a function of the bulk gap size $M_0$ and the exchange coupling parameter $g$ of the Cr-doped TI layers. Both $M_0$ and $g$ change with the Cr doping concentration $x$ in the magnetic TI layers. The calculated Chern number $C$ for points I to V is shown in (**f**). **e**, The calculated phase diagram for the $m = 2$ sample as a function of the thickness $d$ and the gap size $M_0$ of the middle Cr-doped TI layer. The calculated Chern number $C$ for the $m = 2$ samples with $d$ of 0 to 5 nm is shown in (**f**). The color in (**d**) and (**e**) indicates the magnitude of log $(E)$, where $E$ is the calculated energy gap of the $m = 2$ heterostructure sample. **f**, The calculated Chern number $C$ for all points shown in (**d**) and (**e**).

**References:**

1. Haldane, F. D. M. Model for a Quantum Hall-Effect without Landau Levels: Condensed-Matter Realization of the "Parity Anomaly". *Phys. Rev. Lett.* **61**, 2015-2018 (1988).

2. Yu, R., Zhang, W., Zhang, H. J., Zhang, S. C., Dai, X. & Fang, Z. Quantized Anomalous Hall Effect in Magnetic Topological Insulators. *Science* **329**, 61-64 (2010).

3. Chang, C. Z., Zhang, J. S., Feng, X., Shen, J., Zhang, Z. C., Guo, M. H., Li, K., Ou, Y. B., Wei, P., Wang, L. L., Ji, Z. Q., Feng, Y., Ji, S. H., Chen, X., Jia, J. F., Dai, X., Fang, Z., Zhang, S. C., He, K., Wang, Y. Y., Lu, L., Ma, X. C. & Xue, Q. K. Experimental Observation of the Quantum Anomalous Hall Effect in a Magnetic Topological Insulator. *Science* **340**, 167-170 (2013).

4. Chang, C. Z., Zhao, W. W., Kim, D. Y., Zhang, H. J., Assaf, B. A., Heiman, D., Zhang, S. C., Liu, C. X., Chan, M. H. W. & Moodera, J. S. High-Precision Realization of Robust Quantum Anomalous Hall State in a Hard Ferromagnetic Topological Insulator. *Nat. Mater.* **14**, 473-477 (2015).

5. Deng, Y., Yu, Y., Shi, M., Guo, Z., Xu, Z., Wang, J., Chen, X. & Zhang, Y. Quantum anomalous Hall effect in intrinsic magnetic topological insulator. *Science* **367**, 895-900 (2020).

6. Kou, X. F., Guo, S. T., Fan, Y. B., Pan, L., Lang, M. R., Jiang, Y., Shao, Q. M., Nie, T. X., Murata, K., Tang, J. S., Wang, Y., He, L., Lee, T. K., Lee, W. L. & Wang, K. L. Scale-Invariant Quantum Anomalous Hall Effect in Magnetic Topological Insulators beyond the Two-Dimensional Limit. *Phys. Rev. Lett.* **113**, 137201 (2014).

7. Checkelsky, J. G., Yoshimi, R., Tsukazaki, A., Takahashi, K. S., Kozuka, Y., Falson, J., Kawasaki, M. & Tokura, Y. Trajectory of the Anomalous Hall Effect towards the Quantized State in a Ferromagnetic Topological Insulator. *Nat. Phys.* **10**, 731-736 (2014).

8. Mogi, M., Yoshimi, R., Tsukazaki, A., Yasuda, K., Kozuka, Y., Takahashi, K. S., Kawasaki, M. & Tokura, Y. Magnetic Modulation Doping in Topological Insulators toward Higher-Temperature Quantum Anomalous Hall Effect. *Appl. Phys. Lett.* **107**, 182401 (2015).

9. Ou, Y., Liu, C., Jiang, G., Feng, Y., Zhao, D., Wu, W., Wang, X. X., Li, W., Song, C.,

Wang, L. L., Wang, W., Wu, W., Wang, Y., He, K., Ma, X. C. & Xue, Q. K. Enhancing the Quantum Anomalous Hall Effect by Magnetic Codoping in a Topological Insulator. *Adv. Mater.* **30**, 1703062 (2017).

10   Serlin, M., Tschirhart, C. L., Polshyn, H., Zhang, Y., Zhu, J., Watanabe, K., Taniguchi, T., Balents, L. & Young, A. F. Intrinsic quantized anomalous Hall effect in a moiré heterostructure. *Science* **367**, 900-903 (2020).

11   Sharpe, A. L., Fox, E. J., Barnard, A. W., Finney, J., Watanabe, K., Taniguchi, T., Kastner, M. A. & Goldhaber-Gordon, D. Emergent Ferromagnetism near Three-Quarters Filling in Twisted Bilayer Graphene. *Science* **365**, 605-608 (2019).

12   Thouless, D. J., Kohmoto, M., Nightingale, M. P. & Dennijs, M. Quantized Hall Conductance in a Two-Dimensional Periodic Potential. *Phys. Rev. Lett.* **49**, 405-408 (1982).

13   Weng, H. M., Yu, R., Hu, X., Dai, X. & Fang, Z. Quantum Anomalous Hall Effect and Related Topological Electronic States. *Adv. Phys.* **64**, 227-282 (2015).

14   Liu, C. X., Qi, X. L., Dai, X., Fang, Z. & Zhang, S. C. Quantum anomalous Hall effect in $Hg_{1-y}Mn_yTe$ quantum wells. *Phys. Rev. Lett.* **101**, 146802 (2008).

15   Qi, X. L., Hughes, T. L. & Zhang, S. C. Topological Field Theory of Time-Reversal Invariant Insulators. *Phys. Rev. B* **78**, 195424 (2008).

16   Landauer, R. Spatial Variation of Currents and Fields Due to Localized Scatterers in Metallic Conduction. *IBM J. Res. Dev.* **1**, 223-231 (1957).

17   Chang, C. Z., Zhao, W. W., Kim, D. Y., Wei, P., Jain, J. K., Liu, C. X., Chan, M. H. W. & Moodera, J. S. Zero-Field Dissipationless Chiral Edge Transport and the Nature of Dissipation in the Quantum Anomalous Hall State. *Phys. Rev. Lett.* **115**, 057206 (2015).

18   Wang, J., Lian, B. A., Zhang, H. J., Xu, Y. & Zhang, S. C. Quantum Anomalous Hall Effect with Higher Plateaus. *Phys. Rev. Lett.* **111**, 136801 (2013).

19   Fang, C., Gilbert, M. J. & Bernevig, B. A. Large-Chern-Number Quantum Anomalous Hall Effect in Thin-Film Topological Crystalline Insulators. *Phys. Rev. Lett.* **112**, 046801 (2014).

20   Ge, J., Liu, Y., Li, J., Li, H., Luo, T., Wu, Y., Xu, Y. & Wang, J. High-Chern-Number and High-Temperature Quantum Hall Effect without Landau Levels. *Natl. Sci. Rev.* **7**, 1280-

1287 (2020).

21  Chen, G. R., Sharpe, A. L., Fox, E. J., Zhang, Y. H., Wang, S. X., Jiang, L. L., Lyu, B. S., Li, H. Y., Watanabe, K., Taniguchi, T., Shi, Z. W., Senthil, T., Goldhaber-Gordon, D., Zhang, Y. B. & Wang, F. Tunable correlated Chern insulator and ferromagnetism in a moire superlattice. *Nature* **579**, 56-61 (2020).

22  Jiang, H., Qiao, Z. H., Liu, H. W. & Niu, Q. Quantum anomalous Hall effect with tunable Chern number in magnetic topological insulator film. *Phys. Rev. B* **85**, 045445 (2012).

23  Zhang, J. S., Chang, C. Z., Tang, P. Z., Zhang, Z. C., Feng, X., Li, K., Wang, L. L., Chen, X., Liu, C. X., Duan, W. H., He, K., Xue, Q. K., Ma, X. C. & Wang, Y. Y. Topology-Driven Magnetic Quantum Phase Transition in Topological Insulators. *Science* **339**, 1582-1586 (2013).

24  Chang, C. Z., Tang, P. Z., Wang, Y. L., Feng, X., Li, K., Zhang, Z. C., Wang, Y. Y., Wang, L. L., Chen, X., Liu, C. X., Duan, W. H., He, K., Ma, X. C. & Xue, Q. K. Chemical-Potential-Dependent Gap Opening at the Dirac Surface States of $Bi_2Se_3$ Induced by Aggregated Substitutional Cr Atoms. *Phys. Rev. Lett.* **112**, 056801 (2014).

25  Wang, F., Zhang, H., Jiang, J., Zhao, Y.-F., Yu, J., Liu, W., Li, D., Chan, M. H. W., Sun, J., Zhang, Z. & Chang, C.-Z. Chromium-Induced Ferromagnetism with Perpendicular Anisotropy in Topological Crystalline Insulator SnTe (111) Thin Films. *Phys. Rev. B* **97**, 115414 (2018).

26  Burkov, A. A. & Balents, L. Weyl Semimetal in a Topological Insulator Multilayer. *Phys. Rev. Lett.* **107**, 127205 (2011).

27  Belopolski, I., Xu, S. Y., Koirala, N., Liu, C., Bian, G., Strocov, V. N., Chang, G. Q., Neupane, M., Alidoust, N., Sanchez, D., Zheng, H., Brahlek, M., Rogalev, V., Kim, T., Plumb, N. C., Chen, C., Bertran, F., Le Fevre, P., Taleb-Ibrahimi, A., Asensio, M. C., Shi, M., Lin, H., Hoesch, M., Oh, S. & Hasan, M. Z. A novel artificial condensed matter lattice and a new platform for one-dimensional topological phases. *Sci. Adv.* **3**, e1501692 (2017).

28  Jiang, G., Feng, Y., Wu, W., Li, S., Bai, Y., Li, Y., Zhang, Q., Gu, L., Feng, X., Zhang, D., Song, C., Wang, L., Li, W., Ma, X.-C., Xue, Q.-K., Wang, Y. & He, K. Quantum Anomalous Hall Multilayers Grown by Molecular Beam Epitaxy. *Chin. Phys. Lett.* **35**, 076802 (2018).

29  Wang, J., Lian, B., Zhang, H. J. & Zhang, S. C. Anomalous Edge Transport in the Quantum Anomalous Hall State. *Phys. Rev. Lett.* **111**, 086803 (2013).

30  Feng, X., Feng, Y., Wang, J., Ou, Y., Hao, Z., Liu, C., Zhang, Z., Zhang, L., Lin, C., Liao, J., Li, Y., Wang, L. L., Ji, S. H., Chen, X., Ma, X., Zhang, S. C., Wang, Y., He, K. & Xue, Q. K. Thickness Dependence of the Quantum Anomalous Hall Effect in Magnetic Topological Insulator Films. *Adv. Mater.* **28**, 6386-6390 (2016).

31  Liu, C. X., Qi, X. L., Zhang, H. J., Dai, X., Fang, Z. & Zhang, S. C. Model Hamiltonian for topological insulators. *Phys. Rev. B* **82**, 045122 (2010).

32  Zhang, H. J., Liu, C. X., Qi, X. L., Dai, X., Fang, Z. & Zhang, S. C. Topological Insulators in $Bi_2Se_3$, $Bi_2Te_3$ and $Sb_2Te_3$ with a Single Dirac Cone on the Surface. *Nat. Phys.* **5**, 438-442 (2009).

33  Li, R. D., Wang, J., Qi, X. L. & Zhang, S. C. Dynamical axion field in topological magnetic insulators. *Nat. Phys.* **6**, 284-288 (2010).

34  Wang, J., Lian, B. & Zhang, S. C. Dynamical axion field in a magnetic topological insulator superlattice. *Phys. Rev. B* **93**, 045115 (2016).